\documentstyle[12pt,epsf]{article}
\title{Alternative method of generation of Cerenkov radiation or shock
wave}
\author{Amit Halder\\
{\it Department of Applid Physics, }\\
{\it Punjab Engineering College, Chandigarh- 160 012 (India)}}
\begin{document}
\baselineskip 24pt
\maketitle
\begin{abstract}
An alternative method of generation of Cerenkev radiation is proposed
over here with the help of a rotating source and a reflector. The
principle is that, if we focus a narrow beam of light on to source of
light is rotated with certain angular velocity then the light spot on
the surface will move with very high velocity which may exceed the
velocity of light. As a consequence of this we shall observe an effect
very similar to Cerknov radiation.
\end{abstract}
\section*{Introduction}
\begin{figure}
\epsffile{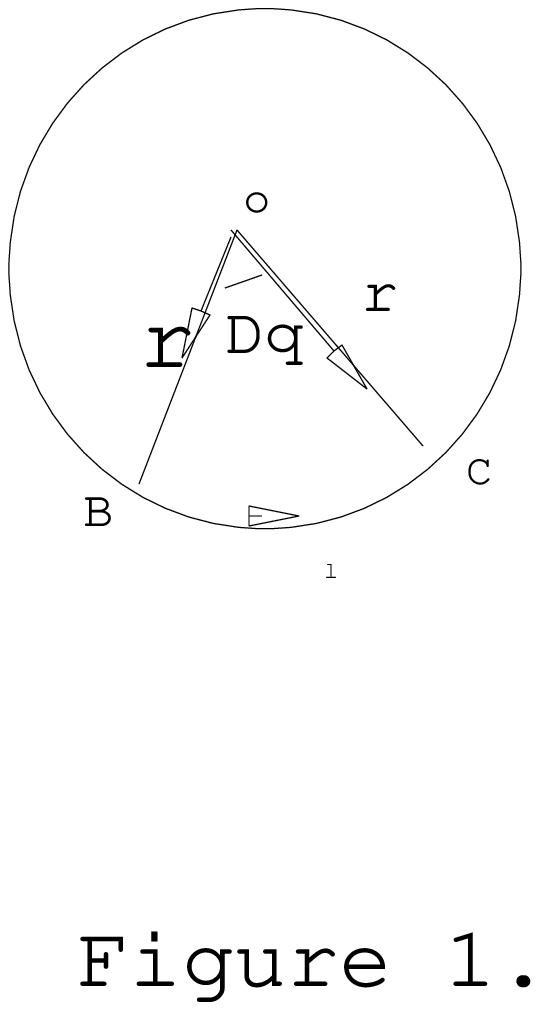}
\caption{}
\label{f1}
\end{figure}

\begin{figure}
\epsffile{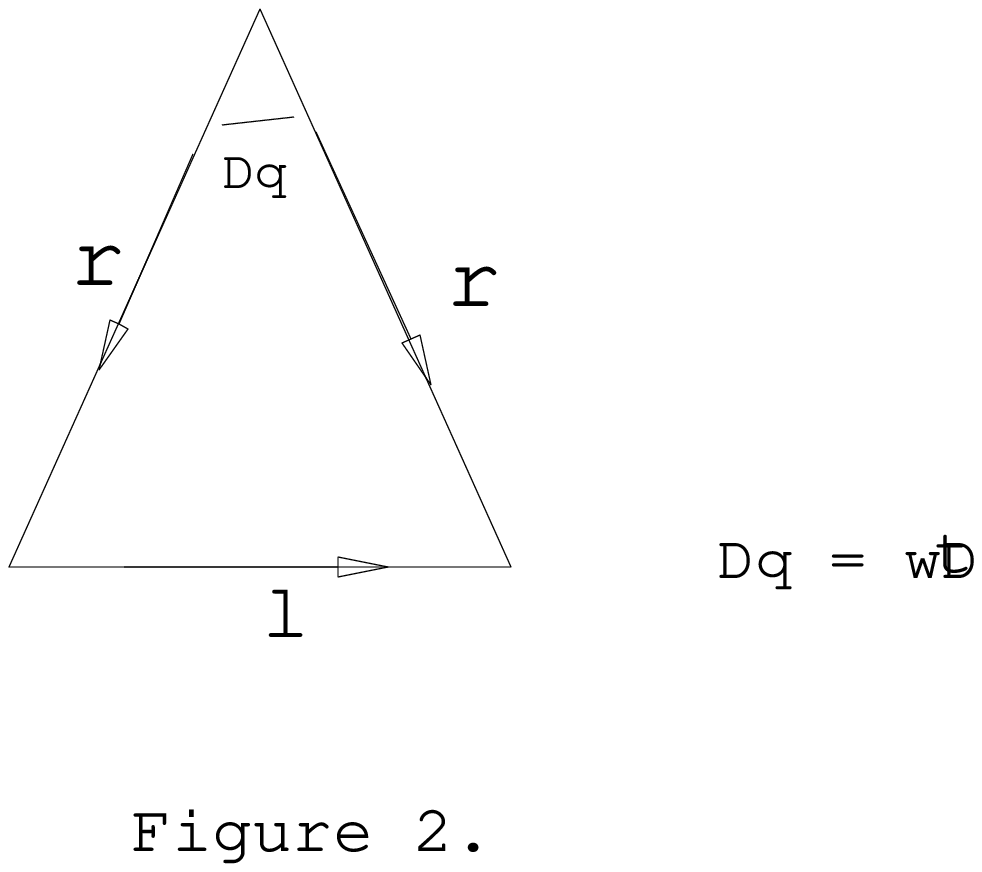}
\caption{}
\label{f2}
\end{figure}

\begin{figure}
\epsffile{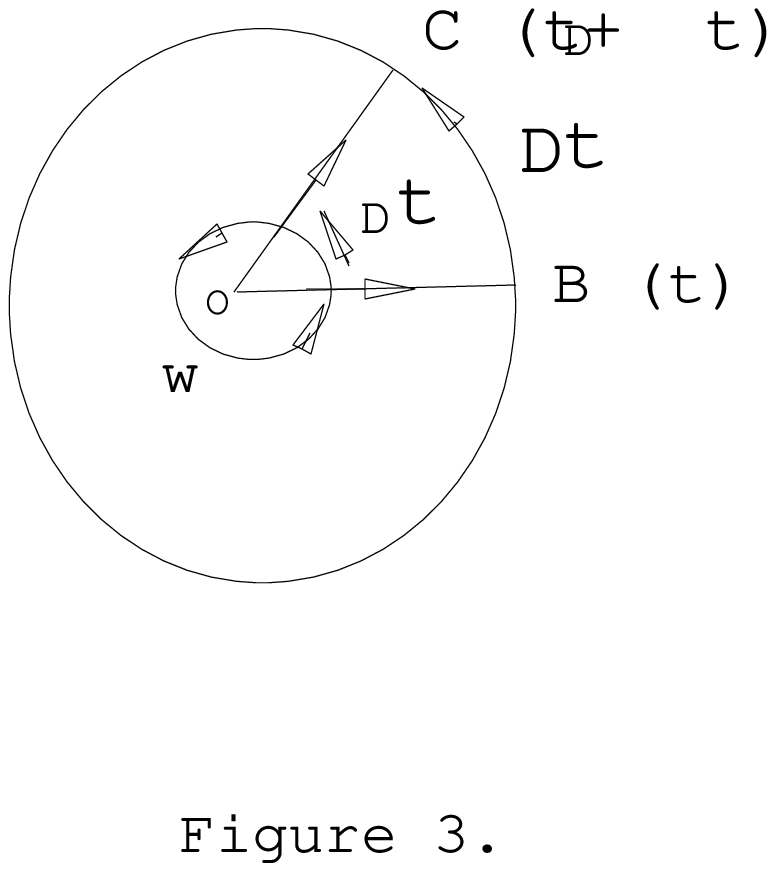}
\caption{}
\label{f3}
\end{figure}

It is an well established fact that a beam of light from a rotating
source may sweep out a distant path faster than the speed of light
\cite{r1,r2}. This can also be established in the following way. A
narrow beam of light of monocharomatic wavelength $\lambda$ from a
source O rotates in the anticlockwise direction with an angular velocity
$\omega$ and describes an angle $\Delta \theta$ in time $\Delta t$
(Fig. \ref{f1}). The radius r of the circle is large (or $\Delta \theta$ be
very small) to consider the path of the light spot BC stright line (Fig.
\ref{f2} )
$$ BC = l = r \Delta \theta $$
$$ \Delta \theta = 1/r $$
$$ \omega =  \Delta \theta / \Delta t $$
The time taken by the light ray to travel from O to B (Fig. \ref{f3})
i.e. the distance r is t. In time $\Delta t$ the light at O is rotated
by an angle $\Delta \theta$ and is in the direction OC. Since, light
takes a time t to travel a distance r, so it will reach the point C
after a time $(t + \Delta t )$. so, $\Delta t$ is the time taken by the
light spot to travel a distance BC = l. Therefore the velocity of the
light spot is $\nu = 1/ \Delta t $
$$ or , \nu = r \Delta \theta / \Delta t (Since, l = r \Delta \theta )$$
$$ or, \nu = r\omega$$
So, the velocity of the light spot = angular velocity of the source
$\times $ radius of the circle on which the light spot moves
\par
\begin{figure}
\epsffile{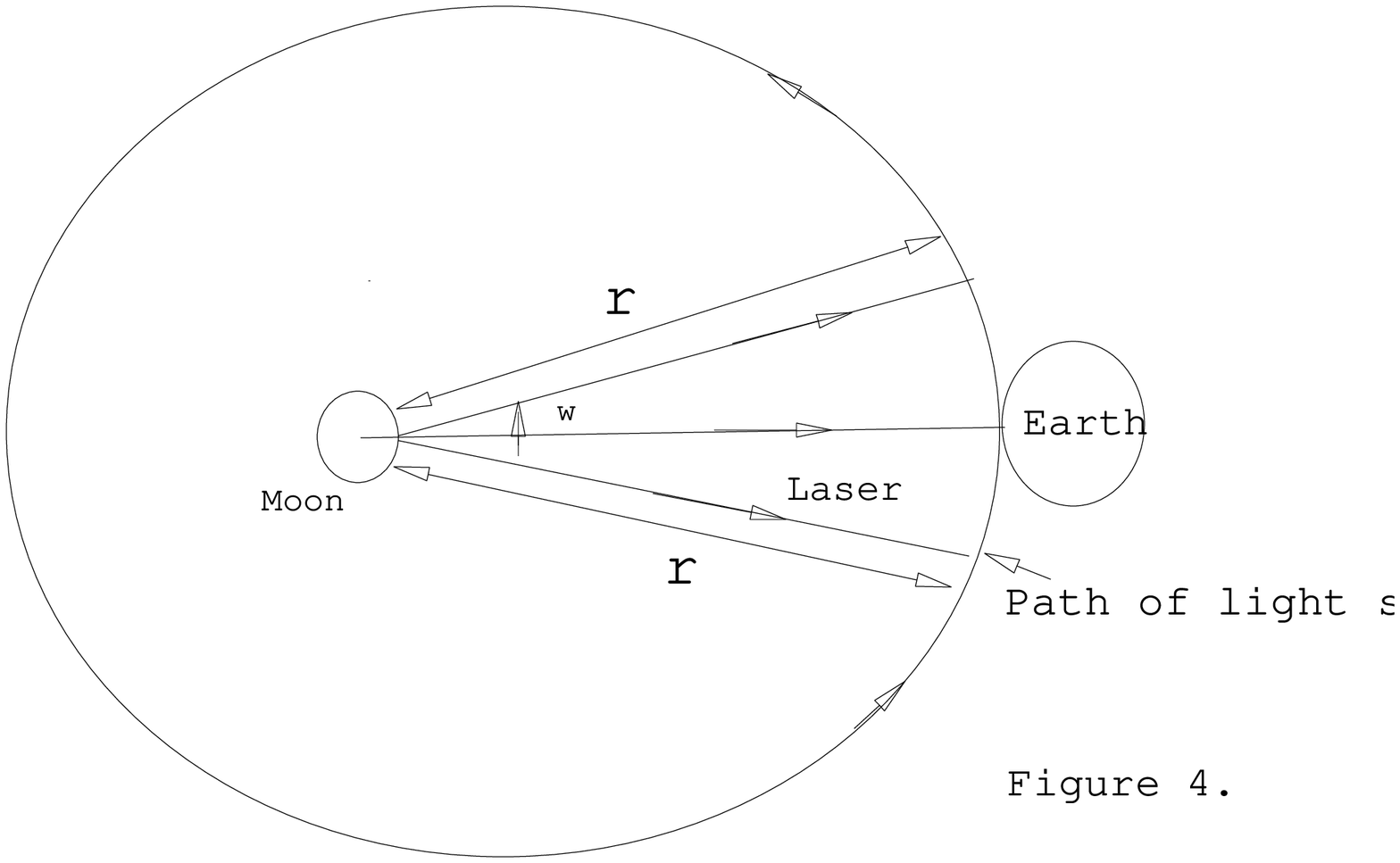}
\caption{}
\label{f4}
\end{figure}

We can make the velocity of the light spot $\nu$ greater than the
velocity of light, either by increasing $\omega$ or by increasing r or
both. As an example (Fig. \ref{f4}) if a beam of laser is rotated at the
surface of moon with one rotation per second (i.e. T = 1 sec, and
$\omega = 2\pi /T = 2\pi /1)$, then the velocity of the light spot which
touches the earth will move with a velocity
$$ \nu = r \omega = r \cdot 2 \pi /T $$
$$ \nu 21.6 \times 10^{10} Con /sec. $$
{\em [taking, $r = 3.37 \times 10^{10} $ cm, the distance between moon and
earth and T = 1 sec. ]}\\
Which is about seven times greater than the velocity of light.
\par
The possiblity that a light-spot can move faster than light is discussed
in references 1 and 2. So, when a spot of light moves faster than the
light itself then we can expect an effect very similar to Cerenkov
radiation \cite{r3}.
\section*{Thought Experiment}
\begin{figure}
\epsffile{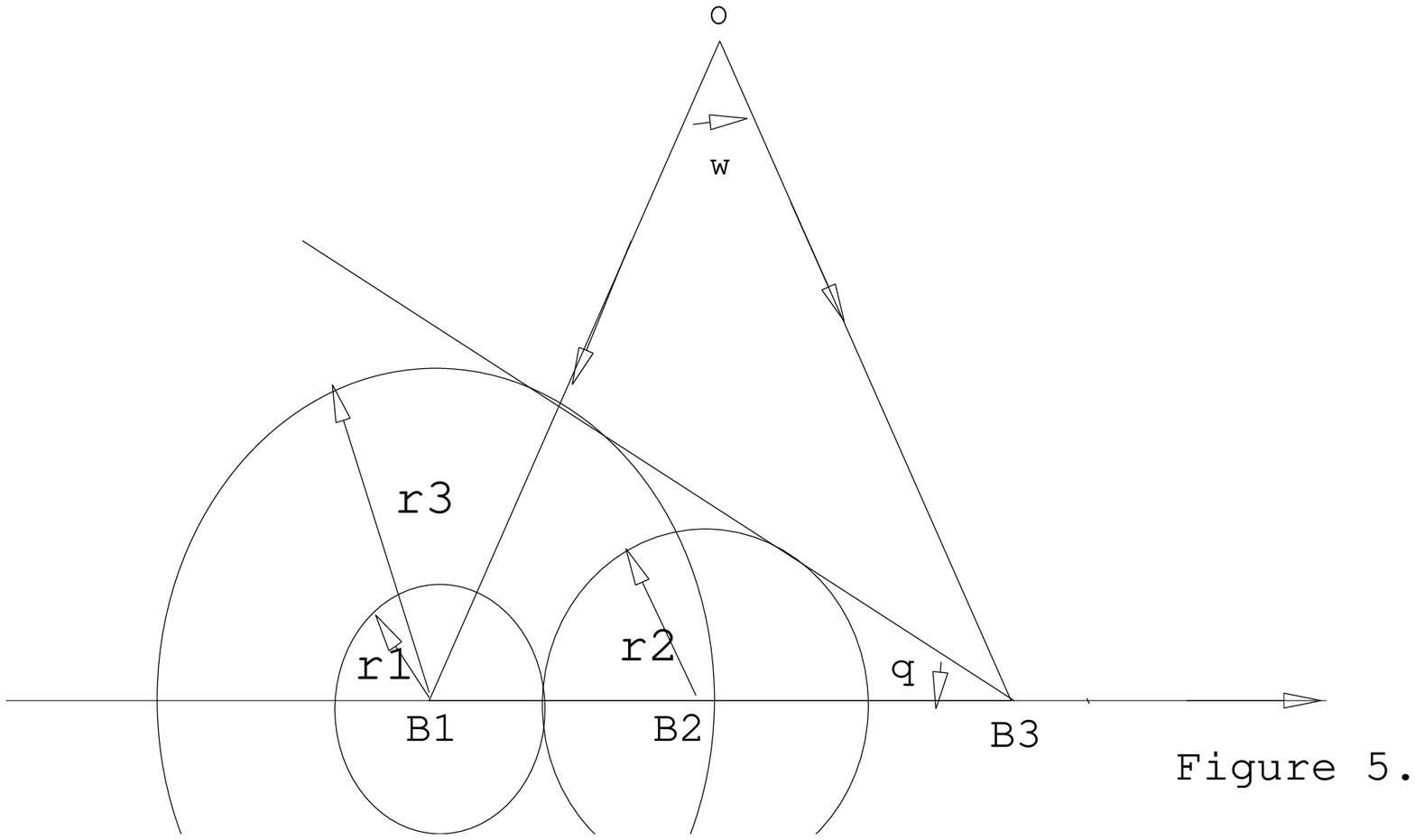}
\caption{}
\label{f5}
\end{figure}

Light from the rotating source O after reflection from  the points like
B will generate reflected or secondary waves (Fig. \ref{f2} and Fig.
\ref{f5}). Since the velocity of this reflected wave is equal to the
velocity of light c, which is less than of the velocity of the
light-spot, so Cerenkov radiation will be generated. This is explained
in the following paragraph.
\par
Suppose at a given moment light from the rotating source O (Fig.
\ref{f5}) falls at $B_1$ and a reflected or secondary wave is generated
at the point $B_1$. Then in the next moment as the light spot (whose
velocity $\nu$ is greater than the velocity of light c ) moves to $B_2$,
the reflected wave from $B_1$ (whose velocity is the velocity of light
c) expands by a radius $r_1$, smaller than the distance the light-spot
moves, then another wave starts from $B_2$. When the light spot moved
still further to $B_3$ and a wave is starting there, the wave from $B_2$
has now expanded to $r_2$ and the one from $B_1$ has expanded to $r_3$.
So, we have a series of wave circles with a common tangent line which
goes through the light spot. The angle $\theta$ (angle between the
common tangent and the line $B_1B_3$) can be calculated easily. In a
given amount of time the light-spot moves a distance $(B_3 - B_1)$,
which is proportional to $\nu$, the velocity of the light spot. In the
mean time the wavefront has moved a distance $r_3$, proportional to the
velocity of light c, therefore, $\sin \theta = c/ \nu.$
\section*{Conclusion}
The effect is very similar to the Cerenkov radiation effect where an
object moving through a medium faster than speed at which the medium
carrier waves will generate waves. When a fast moving charged particle
passes through a block of glass (say), and if the speed of moving
particle is greater than the speed of light in the medium then it will
produce conical waves with its apex at the source. Here, since the
light-spot can move faster  than the speed of light, so it also give
rise to the same effect as Cerenkov radiation. The difference is that
the Cerenkov radiation emitted by the fast moving light-spot can be seen
even in vacuum. If the source O is replaced by a source of sound, then
sock waves will be generated. The source O may be replaced by a beam of
electron, proton etc. then also we will observe the similar type of
effect as discussed above.
\subsection*{Acknowledgement}
The author has got great inspiration from his friend and coleague, Dr.
Ashwani Kumar of the same deaprtment for publishing the article.

\newpage

\end{document}